# Two-Factor Hull-White Model Revisited: Correlation Structure for Two-Factor Interest Rate Model in CVA Calculation

Osamu Tsuchiya[1]

The development of credit valuation adjustment (CVA) (valuation adjustments [XVA]) [Green] has increased the importance of simple interest rate models such as the Hull-White model [Tan14] [Tsuchiya]. This is because the XVA model is an FX hybrid model, and is tractable only when the interest rate part is a simple Gaussian model.

For the XVA calculation of interest rate instruments, de-correlation of the yield curve can be important even for the swap portfolio. Capturing the correlation structure in the two-factor Hull-White model is an integral element of CVA (XVA) modeling. However, the correlation structure in two-factor Hull-White model has not studied enough except for the analysis in [AndersenPiterbarg].

In this study, the correlation structure of the two-factor Hull-White model is analyzed in detail. The correlation structure of co-initial swap rates is investigated using a combination of the approximation formula and Monte-Carlo simulation. The Hull-White model captures the de-correlation of the yield curve only when the parameters (volatilities and mean reversion strength) satisfy certain relationships, making the valuation of XVA by two-factor Hull-White model effective.

1. **Number of factors in the interest rate part of the XVA model**

CVA is given by $\text{CVA} = \int \lambda(t) E[V^+(t)|F_0] dt$ where $\lambda(t)$ is the default intensity and $V(t)$ represents the value of netting set to the counterparty [Green] [Brigo]. Other XVA is also given by the exposure of the netting set (portfolio).In the netting set, various types of derivatives can be observed with several types of currencies. In general, the model needs to be a cross-currency one. Therefore, in the XVA calculation, typically simple models such as the FX hybrid Hull-White model are used [Piterbarg]. The portfolio of netting set can depend on the feature of the counterparty (client). Specifically, some hedge funds can trade the combination of long and short derivatives (payer and receiver swaps)[2].

---

[1] Simplex Inc.

[2] Though the trades with hedge funds are subject to Variation Margin, MVA (Margin Variation Adjustment) and other XVA are necessary. CVA (exposure) is a base to all XVA.

In this article, only the interest rate part of the XVA model is investigated.
there are CMS spread options (yield curve steapners) in the netting set, the exposure depends on correlation of the yield curve.

The other situation, the exposure depends on correlation is when there are payers swaps and recievers swaps in the netting set.

To illustrate the importance of the number of factors, we will show the relationship of CVA calculation and spread option valuation. Let the netting set consist of a payer swap of 20 years with strike $K_1$ (swap1 hereafter) and a receiver swap of 10 years with strike $K_2$ (swap2 hereafter), both quarterly. The exchange of cash flows occurs at times $[T_1, \cdots, T_{80} = 20]$ and $[T_1, \cdots, T_{40} = 10]$, respectively. Let us analyze their expected positive exposure observed at time grid $T_j$.

The positive exposure is given by $\left(A_1(T_j)(S_1(T_j) - K_1) - A_2(T_j)(S_2(T_j) - K_2)\right)^+$

where $S_1(T_j)$ $(A_1(T_j))$ are the par rate (present value of basis point (PVBP)) of the swap which has interest exchanges at $[T_j, \cdots, T_{80} = 20]$ observed at $T_j$ that is

$$A_1(T_j) = \delta \sum_{k=j}^{40} D(T_j, T_k)$$

$$S_1(T_j) = (1 - D(T_j, T_{40}))/A_1(T_j)$$

And $S_2(T_j)$ $(A_2(T_j))$ are the par rate (present value of basis point (PVBP)) of the swap which has interest exchanges at $[T_j, \cdots, T_{20} = 10]$ observed at $T_j$ that is

$$A_2(T_j) = \delta \sum_{k=j}^{20} D(T_j, T_k)$$

$$S_2(T_j) = (1 - D(T_j, T_{20}))/A_1(T_j)$$

The expected positive exposure of time $T_j$ is given as follows:

$$E(T_j) = E\left[\left(A_1(T_j)(S_1(T_j) - K_1) - A_2(T_j)(S_2(T_j) - K_2)\right)^+ | F_0\right]$$

where the numeraire is expected to be normalized in the expectation. Here, because PVBP fluctuates less, freezing it at its time 0 value is a reasonable assumption, and we can approximate it as follows:

$$E(T_j) \sim E\left[\left(A_1(0)S_1(T_j) - A_2(0)S_2(T_j) - A_1(0)K_1 + A_2(0)K_2\right)^+ | F_0\right]$$

This representation of the exposure can be recognized as the value of a spread option (with a different gearing applied to each rate).

To further illustrate the importance of correlation in the XVA calculation, we will analyze a more significant example. Let the netting set consist of a payer forward swap of 5 to 10 years and a receiver forward swap of 10 to 15 years with the same notional and strike. If we use a one-factor model, the yield curve moves are almost exclusively parallel, and the exposures from the first and second swaps cancel each other to a large extent, so that the net exposure observed before five years is close to zero. However, the de-correlation between the shorter and longer tenor rates could lead to a significant exposure. To capture this correlation effect, we need at least two factors in the interest rate part of the XVA model.

The Hull-White model is one of the first-generation interest rate models, and is significantly tractable [Pelsser] [KP]. Its two-factor generalization is also studied [BM] [KK]. Therefore, the two-factor Hull-White model (or its generalization like Cheytte) is widely used for XVA calculation. However, this model faces difficulty to explain de-correlation of the yield curve. [AndersenPiterbarg]. It is known that the Gaussian model cannot adequately explain the de-correlation between forward rates[3]. We investigated this seemingly strange behavior of correlation structure in the two-factor Hull-White model and observed that the model can explain de-correlation of market rates. Thus, this study enables the effective calibration and calculation of CVA (and other XVA) by the two-factor Hull-White model.

2. **Two-factor Hull-White model**

As emphasized in Section 1, the interest rate part of the XVA model should have at least two factors. In this subsection, the two-factor Hull-White model will be summarized[BM][AP].

In the two-factor Hull-White model, the short rate is given by the following:

$$r(t) = x_1(t) + x_2(t) + \phi(t)$$

where

$$x_i(t) = -a_i(t)x_i(t)dt + \sigma_i(t)dW_i(t)$$

$$dW_i(t)dW_j(t) = \rho(t)dt$$

with $(i, j = 1,2)$

Here, by converting the Markov process $x_i(t)$ to a driftless one, we obtain the following:

---

[3] Note that [AP] studied the correlation between instantaneous forward rates, while the only liquid instrument to capture yield curve correlation is the spread option for co-initial swaps.

$$X_i(t) = \exp\left(\int_0^t a_i(u)du\right) x_i(t)$$

$$dX_i(t) = \tilde{\sigma}_i(t)dW_i(t)$$

where $\tilde{\sigma}_i(t) = \exp\left(\left(\int_0^t a_i(u)du\right)\right)\sigma_i(t)$,

discount factors are given by

$$D(t,T) = A(t,T)\exp(-B_1(t,T)X_1(t) - B_2(t,T)X_2(t))$$

where

$$A(t,T) = \frac{D(0,T)}{D(0,t)}\exp\left\{\frac{1}{2}[B_1^2(t,S) - B_1^2(T,S)]\Xi_1(t) + \frac{1}{2}[B_2^2(t,S) - B_2^2(T,S)]\Xi_2(t)\right.$$

$$\left. + [B_1(t,S)B_2(t,S) - B_1(T,S)B_2(T,S)]\Xi_{12}(t)\right\}$$

$$B_i(t,T) = \int_t^T \exp\left(-\int_0^s a_i(v)dv\right)ds$$

$$\Xi_i(T) = \int_0^T \tilde{\sigma}_i^2(u)du$$

$$\Xi_{12}(T) = \int_0^T \tilde{\sigma}_1(u)\tilde{\sigma}_2(u)\rho_{12}(u)du$$

Note that the Hull-White model is a Markov Functional model[KP]. The Markov processes $X_i(t)$ and their variances $\Xi_i(T)$ and covariance $\Xi_{12}(T)$ do not have any financial meaning; rather, the variances and covariances of financial instruments that are traded, such as swaps, have meaning.

### a. Correlation structure of two-factor Hull-White model

We will analyze the correlation between co-initial swap rates in the two-factor Hull-White model in this section. Here, interest exchanges of swaps are on the tenor structure $[0 = T_0, T_1, \cdots, T_{N+1}]$ where all the tenors are $\delta = T_{j+1} - T_j$. The swap rate with interest exchanges at $[T_{n+1}, T_{n+2}, \cdots, T_{m+1}]$ is written as $S_{nm}(t)$.

Here, we will discuss the issues about calibration of the Hull-White model to spread option expired at $T_n$.

For the expiry $T_n$, the calibration instruments are volatilities ($\sigma_{nm}$ and $\sigma_{nl}$), and the correlation ($\rho_{Swap;ml}$) of the two co-initial swap rates $S_{nm}(t)$ and $S_{nl}(t)$. Here $T_m < T_l$ is assumed. For example, $T_m$ is 2 years and $T_l$ is 10 to 20 years.

We assume that the mean reversion parameters are constant and assumed to be $a_1 > a_2$. So:

$$B_1(t,T) = \frac{e^{-a_1 t} - e^{-a_1 T}}{a_1}$$

$$B_2(t,T) = \frac{e^{-a_2 t} - e^{-a_2 T}}{a_2}$$

Now, we will introduce the (terminal) correlation of the Markov process $\rho_M(T_n)$ such that
$$\Xi_{12}(T_n) = \rho_M(T_n)\sqrt{\Xi_1(T_n)\Xi_2(T_n)}$$

We will approximate the swap rate by the forward rate with the same tenor structure as the swap rate, as follows:

$$\frac{D(t,T_{m+1})}{D(t,T_n)} = \frac{1}{(1+\delta S_{nm}(t))^{(T_{m+1}-T_n)/\delta}}$$

In this approximation, the stochastic process of the swap rate is given by:

$$dS_{nm}(T_n) = \left(\frac{1+\delta S_{nm}(0)}{\delta_{nm}}\right)\{-B_1(T_n,T_{m+1})dX_1(t) - B_2(T_n,T_{m+1})dX_2(t)\}$$

where $\delta_{nm} = T_{m+1} - T_n$.

We have an approximate formula for covariance between swap rates, as follows:

$\text{Cov}(S_{nm}(T_n)S_{nl}(T_n)) = \left(\frac{1+\delta S_{nm}(0)}{\delta_{nm}}\right)\left(\frac{1+\delta S_{nl}(0)}{\delta_{nl}}\right)\{B_1(T_n,T_m)B_1(T_n,T_l)\,\Xi_1(T_n) +$

$B_2(T_n,T_m)B_2(T_n,T_l)\Xi_2(T_n) + [B_1(T_n,T_m)B_2(T_n,T_l) + B_2(T_n,T_m)B_1(T_n,T_l)]\,\Xi_{12}(T_n)\}$ (1)

The swap rate (terminal) correlation is given by:

$$\rho_{Swap} = \text{Corr}(S_{nm}(T_n)S_{nl}(T_n)) = \frac{\text{Cov}(S_{nm}(T_n)S_{nl}(T_n))}{\sqrt{\text{Cov}(S_{nm}(T_n)S_{nm}(T_n))\text{Cov}(S_{nl}(T_n)S_{nl}(T_n))}}$$

We will now investigate the correlation structure between the swap rates based on this formula. Although it is an approximation formula, but as will be indicated later, it is accurate enough to investigate the qualitative feature of the model.

When $\rho_M(T_n) = 1$, the correlation between the swap rates is 1, as expected.

On the other hand, in the limit of $\rho_M(T_n) \to -1$ (perfect de-correlation of the Markov process), the situation is different. In this limit, the correlation of swap rates converges to the following:

$$\text{Corr}(S_{nm}(T_n)S_{nl}(T_n))$$
$$\to \frac{\left\{(B_1(T_n,T_m)\sqrt{\Xi_1(T_n)} - B_2(T_n,T_m)\sqrt{\Xi_2(T_n)})(B_1(T_n,T_l)\sqrt{\Xi_1(T_n)} - B_2(T_n,T_l)\sqrt{\Xi_2(T_n)})\right\}}{\sqrt{(B_1(T_n,T_m)\sqrt{\Xi_1(T_n)} - B_2(T_n,T_m)\sqrt{\Xi_2(T_n)})^2(B_1(T_n,T_l)\sqrt{\Xi_1(T_n)} - B_2(T_n,T_l)\sqrt{\Xi_2(T_n)})^2}}$$

In other words, the swap rates correlation can be +1 or -1, depending on the relationship between the parameters (strength of mean reversion and volatilities (variances) of Markov processes).

Actually, because $\frac{B_1(T_n,T_m)}{B_2(T_n,T_m)} > \frac{B_1(T_n,T_l)}{B_2(T_n,T_l)}$, we classify the relationship between the model parameters into the following three cases based on the value of $\sqrt{\frac{\Xi_2(T_n)}{\Xi_1(T_n)}}$:

(I) $\quad \sqrt{\frac{\Xi_2(T_n)}{\Xi_1(T_n)}} > \frac{B_1(T_n,T_m)}{B_2(T_n,T_m)} > \frac{B_1(T_n,T_l)}{B_2(T_n,T_l)}$

(II) $\quad \frac{B_1(T_n,T_m)}{B_2(T_n,T_m)} \geq \sqrt{\frac{\Xi_2(T_n)}{\Xi_1(T_n)}} > \frac{B_1(T_n,T_l)}{B_2(T_n,T_l)}$

(III) $\quad \frac{B_1(T_n,T_m)}{B_2(T_n,T_m)} > \frac{B_1(T_n,T_l)}{B_2(T_n,T_l)} \geq \sqrt{\frac{\Xi_2(T_n)}{\Xi_1(T_n)}}$

The correlation structure is very different according to whether the relationships between the parameters are in case (II), (I), or (III).
In region (II), if the correlation of Markov process $\rho_M(T_n)$ is $-1$, then the swap rate correlation $\rho_{Swap}$ is -1.
On the other hand, if the parameters are in region (I) or (III), when $\rho_M(T_n) \to -1$, the swap rate correlation $\rho_{Swap}$ converges to +1.

When the correlation of the Markov process moves from -1 to +1, the swap rate correlation also moves, as shown in the following graphs. If variances are in region (I) or (III), if the Markov process correlation moves from +1 to -1, the swap rate correlation decreases once and then increases to +1. In the graph, the correlations between analytical approximation and Monte-Carlo simulation are also compared.

[Fig.2]
(horizontal axis=$\rho_M(T_n)$; vertical axis=$\rho_{Swap}$; large dot=Monte-Carlo simulation; and small dot=analytic approximation)

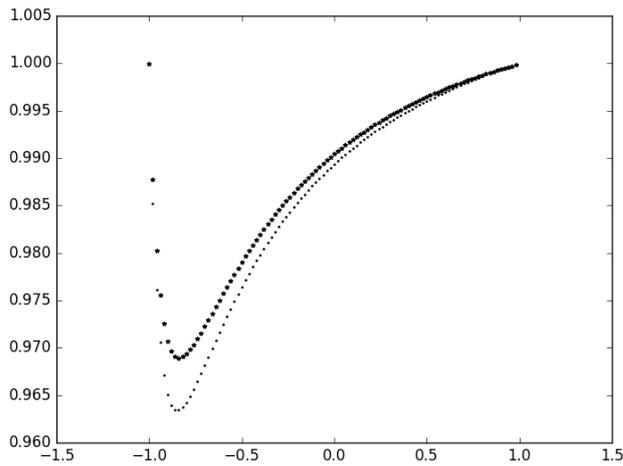

Therefore, if the two variances in the Markov process satisfy conditions (I) or (III), then the model does not explain de-correlation of swap rates sufficiently.

On the other hand, if the variances pair is in region (II), in the limit $\rho_M(T_n) \to -1$, the swap rate correlation converges to -1 and swap rate de-correlation is well explained.

[Fig.3] (horizontal axis=$\rho_M(T_n)$; vertical axis=$\rho_{Swap}$; large dot=Monte-Carlo simulation; and small dot=analytic approximation)

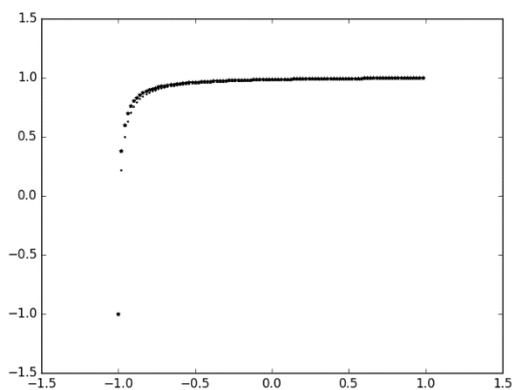

As apparent from the abovementioned conditions, when the length of the short swap is fixed, then as the length of the long swap increases the condition the parameters satisfy moves from region (I) to (II). In other words, as the length of the longer swap increases, the higher is the decrease in the swap rate correlation. This observation matches the intuition.

Note that the correlation between co-initial swap rates is large (typically over 70%). The correlation between Markov processes is calibrated in this area.

[Fig.4]
(Horizontal axis denotes the lengths of longer swap rate $T_l$ with $T_n$ and $T_m$ are fixed.)

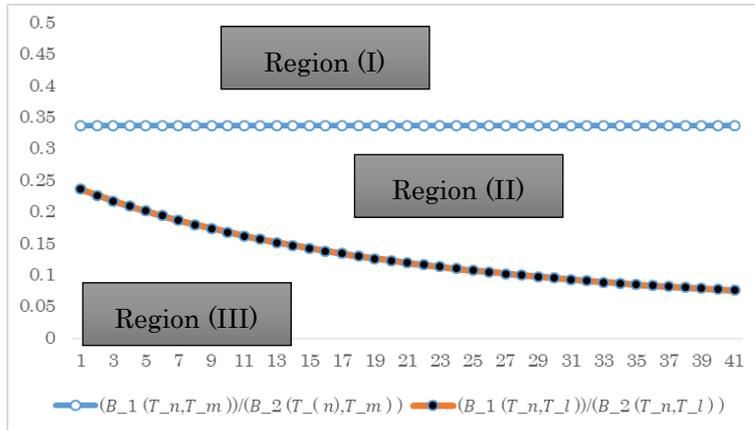

Note that for the given $\frac{B_1(T_n,T_m)}{B_2(T_n,T_m)} \geq \sqrt{\frac{\Xi_2(T_n)}{\Xi_1(T_n)}}$, as maturity of long swap $T_l$ moves from $T_m$ to $\infty$, because $\frac{B_1(T_n,T_l)}{B_2(T_n,T_l)}$ is a decreasing function, the situation moves from region (III) to (II) and the swap rate correlation decreases as expected.

*i. Analysis in Monte-Carlo simulation*

Now we will analyze the correlation structure of swap rates via Monte-Carlo simulation. In the graphs below, scatter plots of short and long swap rates are depicted by gradually changing the correlation of the Markov process.

Note that in the scatter plot in [Fig.5] to [Fig.10], horizontal axis represent shorter swap rate and vertical axis represent longer swap rate.

[Fig.5][4]

($\rho_M(T_n)$=0; $\rho_{Swap}$=0.987315037)

---

[4] Note that in the [Fig.5] to [Fig.10], the calculation is done with ; $\sqrt{\Xi_1(T_n)}$ =0.02; $\sqrt{\Xi_2(T_n)}$=0.3*$\sqrt{\Xi_1(T_n)}$; NbPaths=1000.

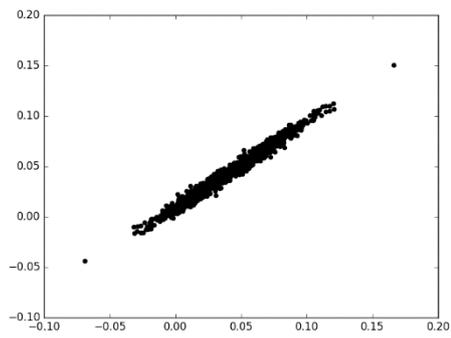

[Fig.6]
($\rho_M(T_n)$=-0.8; $\rho_{Swap}$=0.89250204)

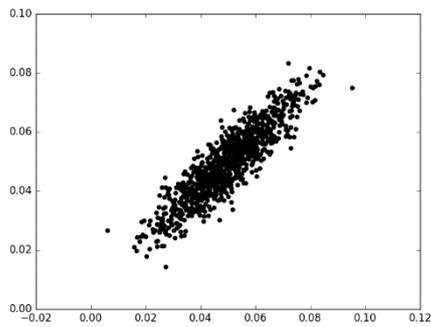

[Fig.7]
(Rho=-0.9; $\rho_{Swap}$=0.78824838)

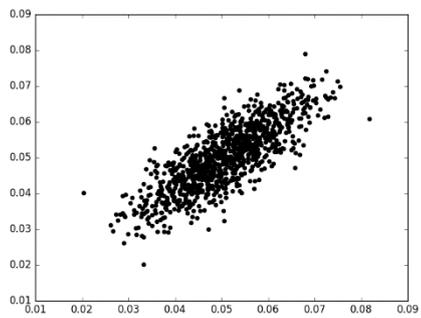

[Fig.8]
($\rho_M(T_n)$=-0.95; $\rho_{Swap}$=0.61559151)

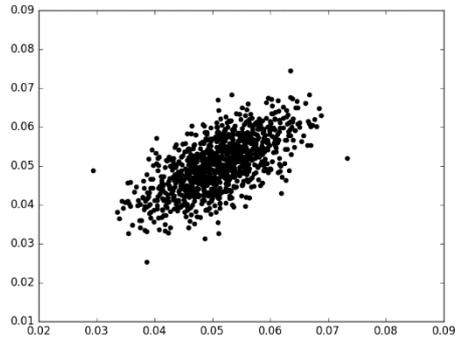

[Fig.9]
($\rho_M(T_n)$=-0.99; $\rho_{Swap}$=-0.054115682)

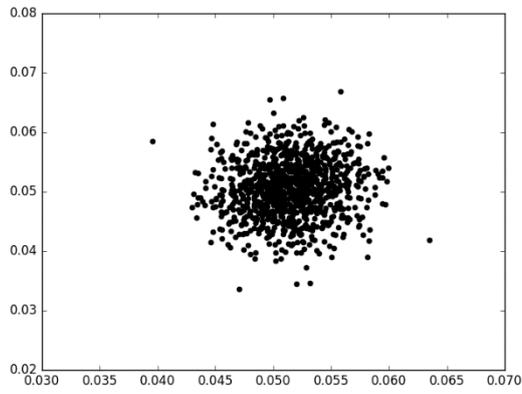

[Fig.10]
($\rho_M(T_n)$=-0.999; $\rho_{Swap}$=-0.820799258)

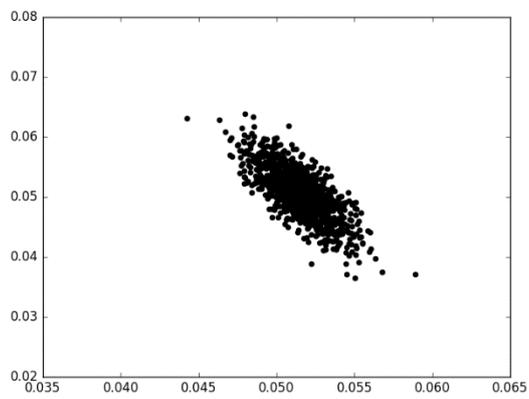

The above analysis shows that if the parameters are in region (II), then a decrease in the correlation between the Markov process appropriately explains the de-correlation between the co-initial swap rates.

At the first glance, it looks strange that even when parameters (or length of longer swap) change continuously, the convergence point of the correlation shifts from –1 to +1, but actually it is not.

In the Monte-Carlo simulation, in the limit of $\rho_M(T_n) \to -1$, the scatter plot between the short swap rate $S_{nm}(T_n)$ and long swap rate $S_{nl}(T_n)$ changes as the parameters move from region (III) to (II) as shown in the graph below.

[Fig.11][5]
(In region (III), $\sqrt{\Xi_2(T_n)/\Xi_1(T_n)}$=0.23)
(Swap Rate Correlation ~ +1)

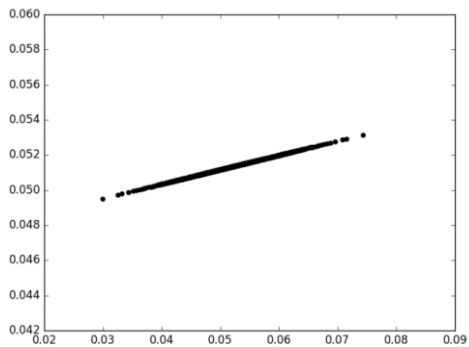

[Fig.12]
(In the boundary between regions (III) and (II), $\sqrt{\Xi_2(T_n)/\Xi_1(T_n)}$=0.24)
(Swap Rate Correlation boundary between +1 and -1)

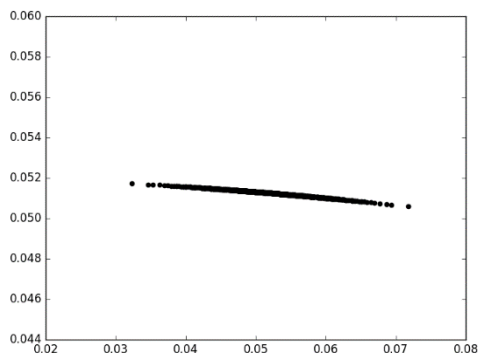

---

[5] Note that [Fig.11], [Fig.12],[Fig.13] is calculated with $\sqrt{\Xi_1(T_n)}$=0.02, $\rho_M(T_n)$=-0.99999999; NbPaths=1000.

[Fig.13]
(In region (II), $\sqrt{\Xi_2(T_n)/\Xi_1(T_n)}$=0.25)
(Swap Rate Correlation~ －1)

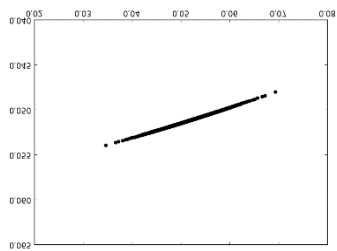

In the above figures, when the parameters are in region [III], the scatter plot has a positive slope, and as they reach the boundary of region [II], the slope becomes close to flat. Finally, when they reach region [II], the slope becomes negative. The movement of the scatter plot is continuous but the correlation changes from +1 to -1 discontinuously.

## 3  Exposure Calculation

In the previous section, the condition that the de-correlation is expressed in the two-factor Hull-White model is given. In this section, the exposure of the netting set which have both payer and receiver swap will be calculated.
In this example, the (expected positive) exposure of the netting set which consists of payer swap and receiver swap is calculated with changing the Hull-White model correlation.
As is seen in the [Fig 14], when the parameters are in the region (II), when the parameters are in the region (II), when the model correlation move to negative, the exposure increase significantly, where when the parameters are in the region (III), the exposure does not move significantly. This shows that to capture the de-correlation effect enough, the model should be calibrated with the parameters are in the region (II).

[Fig 14]
(The exposure calculated at observation time 10 years. The payer swap has notional 2,000,000 USD and the final maturity is 20 years. The receiver swap has notional of 1,000,000 USD and the final maturity is 12 years. The model is calibrated to European swaption with expiry is 10 years and swap tenor is 10 years. The swaption normal volatility is 1%.)

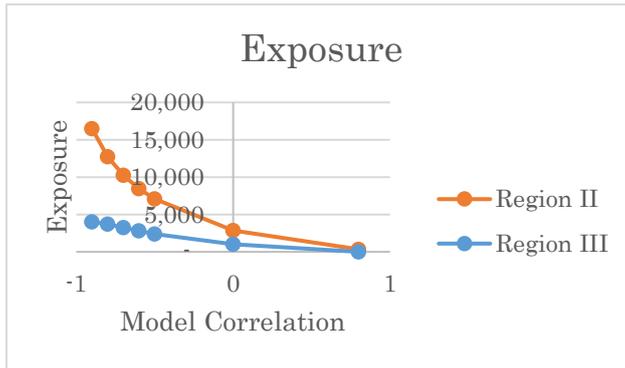

## 4. Conclusion

The two-factor Hull-White model has three regions for the relationships that the parameters satisfy. The qualitative feature of the correlation structure is different in each region. The model can explain correlation well only when the parameters are in one of these regions. In this case, the model can be calibrated to the spread option market and capture the de-correlation The analysis in this study suggests that the valuation of CVA can be effective.

The opinions expressed in this work are solely those of the authors and do not represent in any way those of their current and past employers.

## A.1  Bibliography